\documentclass[epj]{svjour}
\usepackage{graphicx}
\usepackage{amsmath}

\begin{document}
\title{Large-deviation properties of largest component for random graphs}
\author{A. K. Hartmann\inst{1}}
\institute{
Institut f\"ur Physik,
Carl von Ossietzky Universit\"at Oldenburg,
26111 Oldenburg, Germany\\
 \email{a.hartmann@uni-oldenburg.de} }

\date{Received: date / Revised version: date}

\abstract{
Distributions of the size of the largest
component, in particular the large-deviation tail,
 are studied  numerically for two graph ensembles,
for Erd\H{o}s-R\'enyi random graphs 
with finite connectivity and for two-dimensional bond percolation. Probabilities as 
small as $10^{-180}$ are accessed 
 using an artificial finite-temperature (Boltzmann)
ensemble. The distributions for  the Erd\H{o}s-R\'enyi ensemble
agree well with previously obtained analytical
results. The results for the percolation problem, where no analytical results
are available, are qualitatively similar, but the shapes of the distributions
are somehow different and the finite-size
corrections are sometimes much larger. Furthermore, for both problems,
a first-order phase transition
at low temperatures $T$ within the artificial ensemble is found in the
percolating regime, respectively.
}

\maketitle

%
%
\section{Introduction}
\label{sec:intro}

For many problems in science and in statistics, the large deviation
properties play an important role \cite{denHollander2000,dembo2010}.
Only for few cases analytical results can be obtained.
Thus, most problems have to be studied by numerical simulations 
\cite{practical_guide2009}, in particular by Monte Carlo (MC) techniques
\cite{newman1999,landau2000}. Classically, MC simulation have
been applied to random systems in the following way:
For a finite set of independently drawn quenched random instances 
regular or large-deviation properties of these instances have been
calculated using importance-sampling MC simulations. 
Only recently it has been noticed  that by introducing an artificial
sampling temperature also the
large-deviation properties with respect to the quenched random ensemble
can be obtained \cite{align2002}. This corresponds somehow to an annealed 
average, but the results are re-weighted in a way that the results for 
the original quenched ensemble are obtained. In this way, the
large-deviation properties of the distribution of alignment
scores for protein comparison was studied 
\cite{align2002,align_long2007,newberg2008}, 
which is of importance to calculate
the significance of results of protein-data-base queries \cite{durbin2006}.

Motivated by these results, similar approaches have been applied to other
problems like the distribution of the number of components of
Erd\H{o}s-R\'enyi (ER) random graphs \cite{rare-graphs2004},
the partition function of Potts models \cite{partition2005},
the distribution of ground-state energies of spin glasses
\cite{pe_sk2006} and of directed polymers in random media \cite{monthus2006},
the distribution of Lee–-Yang zeros for spin glasses \cite{matsuda2008},
the distribution of success probabilities of error-correcting codes
\cite{iba2008}, the distribution of free energies of RNA secondary
structures \cite{rnaFreeDistr2010}, and some large-deviation properties
of random matrices \cite{driscoll2007,saito2010}.

Interestingly,
so far no comparison between numerical and mathematically exact results for
the full support of a distribution involving a large-deviation tail
has been performed to the knowledge
of the author. In few cases, numerical results had been compared
to rigorous results \cite{monthus2006}, but only 
near the peak of the distribution, where finite-size effects are 
often very small.
In another case \cite{rare-graphs2004}, the numerical and analytical
distributions were compared on the full support, but the result
was obtained using a non-rigorous statistical mechanics approach.
In this work, the numerical large-deviation approach is applied to obtain 
the complete distribution of the size $S$ of the largest component of 
ER random graphs. In this case also mathematically exact results for the 
leading order large-deviation rate function are available for 
arbitrary values of the finite connectivity $c$. This allows
for a comprehensive comparison and an estimation of the strength of 
finite-size effects.
Furthermore, in this paper results are obtained for the two-dimensional
(2d) percolation
problem, where no analytic results are available
for the distribution of the largest-component size $S$. For both models
also the dependence of $S$ on the artificial
sampling temperature $T$ is studied. First-order phase transitions are
found for both graph ensembles in the percolating regime.

The two graph ensembles are defined as follows. In both cases,
each graph $G=(V,E)$ consists of $N$ nodes $i \in V$ and undirected
edges $\{i,j\}\in E \subset V^{(2)}$. For ER random graphs 
\cite{erdoes1960}, each possible
edge $\{i,j\}$ is present with probability $c/N$. 
Hence, the average degree (connectivity) is $c$.
For 2d percolation,
the graph is embedded in a two-dimensional square lattice of size $N=L\times L$
with periodic boundary conditions in both directions, i.e., in a torus. 
Each node can be
connected by edges to its four nearest neighbours, each edge is present with 
probabilty $p$. Hence, the average deegree is $4p$.

Two nodes $i,j$ are called \emph{connected} if there exist a \emph{path}
of disjoint edges $\{i_0,i_1\},\{i_1,i_2\},\ldots, \{i_{l-1},i_l\}$
such that $i=i_0$ and $j=i_l$. The maximum-size subsets $C\subset V$ of nodes,
such that all pairs $i,j\in C$ are connected are called the (connected)
\emph{components} of a graph. The size of the largest component of a graph
is denoted here by $S$. Via the random-graph enembles, a probability
distribution $P(S)$ for the size of the largest component 
and the corresponding
probability $P(s)$ for relative sizes $s=S/N$ are defined.
The probabilities $P(s)$ for values of $s$
different from the typical size
are exponentiall small in $N$. Hence, one uses the concept of the
large-deviation \emph{rate function} \cite{denHollander2000} by writing

\begin{equation}
P(s) = e^{-N\Phi(s)+o(N)}\quad (N\to\infty)
\end{equation}

This leading-order behavior of the large-deviation rate
function  $\Phi_{\rm ER}(s,c)$
 for ER random graphs with connectivity $c$ is known exactly  \cite{biskup2007}
and given by the following set of equations

\begin{eqnarray}
\tilde S(s) & = & s \log s + (1-s)\log(1-s)\nonumber\\
\pi_1(\alpha) & = & 1- e^{-\alpha} \nonumber \\
\Psi(\alpha) & = & \left( \log \alpha - 0.5[\alpha-1/\alpha]\right)
\wedge 0\nonumber \\
\Phi_{\rm ER}(s,c) & = 
& \tilde S(s)-s\log \pi_1(cs) - (1-s)\log \left(1-\pi_1(cs)\right)
\nonumber \\
& & -(1-s)\Psi\left(c(1-s)\right)\;,
\label{eq:rate_fct_analytic}
\end{eqnarray}
where the expression $g(\alpha)\wedge 0$ results in 0 
if $g(\alpha)>0$ and in $g(\alpha)$ else. 

For percolation in finite dimensions similar results are not available, to 
the knowledge of the author.
There are only analytical results for the distribution of finite
(non-percolating) components \cite{alexander1990}.

The paper is organised as follows. In the second section, the numerical
simulation technique and the corresponding re-weighting approach are
explained. In the third section, the results are displayed, first for the
ER random graph ensemble, next for the 2d percolation problem.
Finally, a summary and an outlook are given.
A concise summary of this paper is available at the \emph{papercore} web page
\cite{papercore}.

%
%
\section{Simulation and reweighting method} 
\label{sec:method}

To determine the distribution $P(S)$ for any measurable quantity $S$,
here denoting the largest
component for an ensemble of graphs, \emph{simple sampling} is
straightforward: One generates a certain number $K$ of graph samples
and obtains $S(G)$ for each sample $G$.
This means each graph $G$ will appear with its natural ensemble 
probability $Q(G)$.
The probability to measure a value of $S$ is given by
\begin{equation}
P(S) = \sum_{G} Q(G)\delta_{S(G),S} \label{eq:PS}
\end{equation}
Therefore, by calculating a histogram of the values for $S$, a good
estimation for $P(S)$ is obtained.
Nevertheless, $P(S)$ can only be measured in a regime where $P(S)$
is relatively large, about $P(S)>1/K$. Unfortunately, the distribution
decreases exponentially fast in the system size $N$ when moving
away from its typical (peak) value.
This means, even for moderate system
sizes $N$, the distribution will be unknown on almost its complete support.

To estimate $P(S)$ for a much larger range,  even possibly on the 
full support of $P(S)$, where probabilities smaller
than $10^{-100}$ may appear,
a different approach is used \cite{align2002}. 
For self-containedness, the method is outlined subsequently.
The basic idea is to use an additional
Boltzmann factor $\exp(-S(G)/T)$, $T$ being a ``temperature''
parameter,  in the following manner:
A standard Markov-chain MC simulation \cite{newman1999,landau2000}
is performed, where in each step
$t$ from the current graph $G(t)$ a candidate graph $G^*$ is created:
A node $i$ of the current graph is selected randomly, 
with uniform weight $1/N$,
and all adjacent edges are deleted. For all feasible edges $\{i,j\}$, the edge
is added with a weight corresponding to the natural weight $Q(G)$, i.e.,
with probability $c/N$ (ER random graph) or with probability $p$
(percolation), respectively. For the candidate
graph, the size $S(G^*)$ of the largest component is calculated. Finally,
the candidate graph is \emph{accepted}, ($G(t+1)=G^*$) 
with the Metropolis probability
\begin{equation}
p_{\rm Met} = \min\left\{1,e^{-[S(G^*)-S(G(t))]/T}\right\}\,.
\end{equation}
Otherwise the current graph is kept ($G(t+1)=G(t)$).  By construction,
the algorithm fulfills detailed balance. Clearly the algorithm is also
ergodic, since within $N$ steps, each possible graph may be constructed. Thus,
in the limit of infinite long Markov chains,
the distribution of graphs will follow the probability
\begin{equation}
q_T(G) = \frac{1}{Z(T)} Q(G)e^{-S(G)/T}\,, \label{eq:qT}
\end{equation}
where $Z(T)$ is the a priori unknown normalisation factor.

The distribution for $S$ at temperature $T$ is given by
\begin{eqnarray}
P_T(S) & = &\sum_{G} q_T(G) \delta_{S(G),S} \nonumber\\
       & \stackrel{(\ref{eq:qT})}{=} & 
        \sum_{G} Q(G)e^{-S(G)/T} \delta_{S(G),S} \nonumber \\
       & = & \frac{e^{-S/T}}{Z(T)} \sum_{G} Q(G) \delta_{S(G),S}
               \nonumber \\
       &  \stackrel{(\ref{eq:PS})}{=} & 
           \frac{e^{-S/T}}{Z(T)} P(S) \nonumber\\
\Rightarrow \quad P(S) & = & e^{S/T} Z(T) P_T(S) \label{eq:rescaling}
\end{eqnarray}
Hence, the target distribution $P(S)$ can be estimated, up to a normalisation
constant $Z(T)$, from sampling at finite temperature $T$. For each
temperature, a specific range of the distribution $P(S)$ will be sampled:
Using a positive temperature allows to sample the region of
a distribution left to its peak (values
smaller than the typical value), 
while negative temperatures are used to access the right tail.
Temperatures of large absolute value will cause a sampling of the 
distribution close to its typical value, while temperatures of small 
absolute value
are used to access the tails of the distribution. Hence, by choosing a
suitable set of temperatures, $P(S)$ can be measured over a large
range, possibly on its full support.

The normalisation constants $Z(T)$ can easily be obtained by including
a histogram obtained from simple sampling, which corresponds
to temperature $T=\pm\infty$, which means $Z\approx 1$ 
(within numerical accuracy). Using suitably chosen
temperatures $T_{+1}$, $T_{-1}$, one measures histograms which overlap 
with the simple sampling histogram on its left and right border,
respectively. Then the corresponding normalisation
constants $Z(T_{\pm 1})$ can be obtained by the requirement that
after rescaling the histograms according to 
(\ref{eq:rescaling}), they must agree
in the overlapping regions with the simple sampling histogram within
error bars. This means, the histograms are ``glued'' together. In the same
manner, the range of covered 
$S$ values can be extended iteratively to the left and to
the right by choosing additional suitable temperatures 
$T_{\pm 2}, T_{\pm 3}, \ldots$ and gluing the 
resulting histograms one to the other. A pedagogical explanation
 and examples of 
this procedure can be found in Ref.\ \cite{align_book}.

\begin{figure}[t!]
  \centering
  \includegraphics[clip,width=0.45\textwidth]{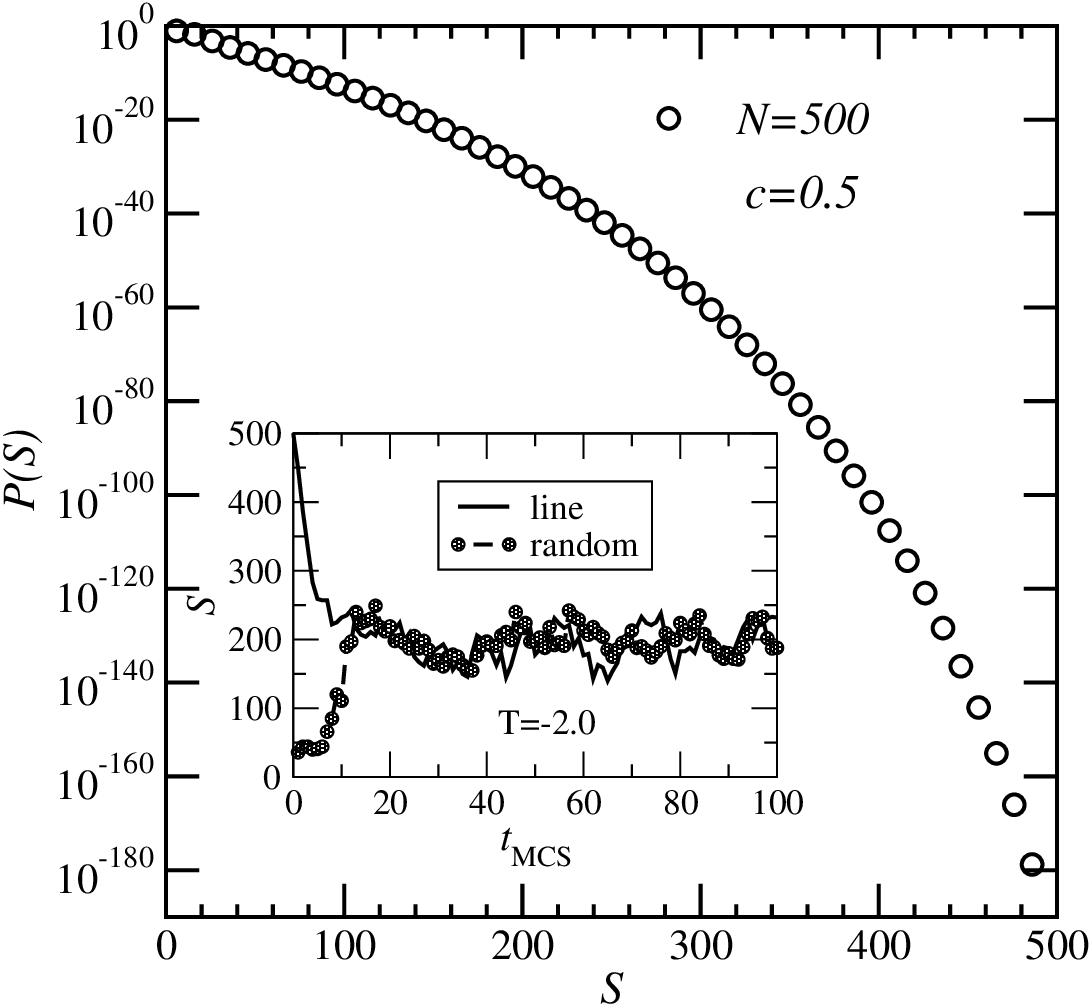}
  \caption{
    \label{fig:equil_distr}
     Distribution of the size $S$ of the largest component for 
Erd\H{o}s-R\'enyi random graphs of size $N=500$ at connectivity $c=0.5$.
In this and all other plots, error bars are of symbol size or smaller
if not explicitly shown.
The inset shows the size of the largest component as function of the number 
$t_{\rm MCS}$ of Monte Carlo sweeps for the same type of graphs at 
temperature $T=-2$. Two different starting conditions are displayed: either
a random graph (leading to a typical components size around $30$) or a graph
consisting of a line $(S=500)$ were used.
}
\end{figure}

In order to obtain the correct result, the MC simulations must be
equilibrated. For the case of the distribution of the size of the largest
component, this is very easy to verify:
The equilibration of the simulation can be monitored by starting
with two different initial graphs, respectively: 
\begin{itemize}
\item Either an unbiased 
random graph is taken, which means that the largest component
is of typical size. In the inset of  Fig.\ \ref{fig:equil_distr}
 the evolution of $S$ as a function of the number
$t_{\rm MCS}=t/N$ of Monte Carlo sweeps is shown for Erd\H{o}s-R\'eny random
graphs with $N=500$ nodes, connectivity $c=0.5$ at temperature $T=2$.
As one can see, $S(t_{\rm MCS})$ moves quickly away from the typical
size which is around $S=30$ towards a values around $S=200$. This shows that 
easily different parts of the distribution can be addressed.
The result of a second run with a negative temperature
is shown in the same inset.
In this case an initial graph was used which consists of a 
single line of nodes, i.e., in particular the graph is fully connected 
leading to $S=N$.

\item Alternatively, if the temperature is positive, one
can start with an empty graph ($S=1$). In any case,
for the two different initial conditions,
the evolution of $S(t_{\rm MCS})$  will approach from two different
extremes, which allows for a simple equilibration test:
equilibration is achieved if
the measured values of $S$ agree within the range of fluctuations.
 Only data was used in this work,
where equilibration was achieved within 200 Monte Carlo steps.
\end{itemize}

The resulting distribution for ER random graphs ($C=2,N=500$) is shown
in the main plot of  Fig.\ \ref{fig:equil_distr}. As one can see, the
distribution can be measured over its full support such that
probabilities as small as $10^{-180}$ are accessible.

Note that in principle one can also use a Wang-Landau approach \cite{wang2001}
or similar approaches 
to obtain the distribution $P(S)$ without the need to perform independent
simulations at different values for the temperatures. 
Nevertheless, the author has performed
tests for ER random graphs and experienced problems by using the 
Wang-Landau approach, because
the sampled distributions tend to stay in a limited fraction of
the values of interest. Using the finite-temperature
approach it is much easier to guide the simulations
to the regions of interest, e.g., where data is missing
using the so-far-obtained data, and to monitor the equilibration process.
Furthermore, the behavior of $S$ as a function of $T$ appears to be
of interest on its own, see next section.

%
%
\section{Results}
\label{sec:results}

ER random graphs of size $N=500$ and 2d percolation problems with lateral
size $L=32$ ($N=1024$ sites) 
were studied. In few cases, additional system sizes were 
considered to estimate the strength of finite-size effects, see below.
For each problem, the model was studied right
at the percolation transition, for one point in the non-percolating
regime, and for one point in the percolating regime.
The temperature ranges used for the different cases are shown in Tab.\
\ref{tab:parameters}. Note that, depending on the position of the peak of the 
size distribution, sometimes positive, sometimes negative and sometimes
both types of temperatures had to be used.
For the systems listed in the table, equilibration was always achieved
within the first 200 MCS. In general, studying significantly 
larger sizes or going deeper
into the percolation regime makes the equilibration much more difficult.
 After equilibration, data was collected for 9800
MCS, in some cases, to improve statistics, for about $10^6$ MCS. In the two
subsequent subsections, the results for ER random graphs and for
2d percolation are presented, respectively.

\begin{table}
\begin{center}
\begin{tabular}{l|lll}
system & $T_1$ & $T_2$ & $N_T$ \\\hline
ER $c=0.5$ & -5 & -0.4 & 14 \\
ER $c=1.0$ & -7.0 & -0.6 & 9\\
ER $c=2.0$ & -2.0 & 10.0 & 6\\
perc $p=0.3$ & -10.0 & -0.6 & 13\\
perc $p=0.5$ & -5.00 & 50.0 & 13\\
perc $p=0.6$ & 0.60 & 30.0 & 14\\
\end{tabular}
\caption{Parameters used to determine the distributions $P(S)$ for the
different models. $T_1$ is the minimum and $T_2$ the maxi\-mum 
temperature used.
$N_T$ denotes the number of different temperature values. Note that for
determinining the average value $\overline{S}(T)$, see Figs.\ \ref{fig:STER}
and \ref{fig:STdistrPerc}, usually a higher number of temperatures was used.
\label{tab:parameters}}
\end{center}
\end{table}

\subsection{ER graphs}

For the numerical simulations, first the case of ER random graphs is treated, 
because the analytical
result  (\ref{eq:rate_fct_analytic}) can be used for comparison. This
allows to assess the quality of the method and to get an impression of
influence of the non-leading finite-size corrections.

\begin{figure}[ht]
  \centering
  \includegraphics[clip,width=0.45\textwidth]{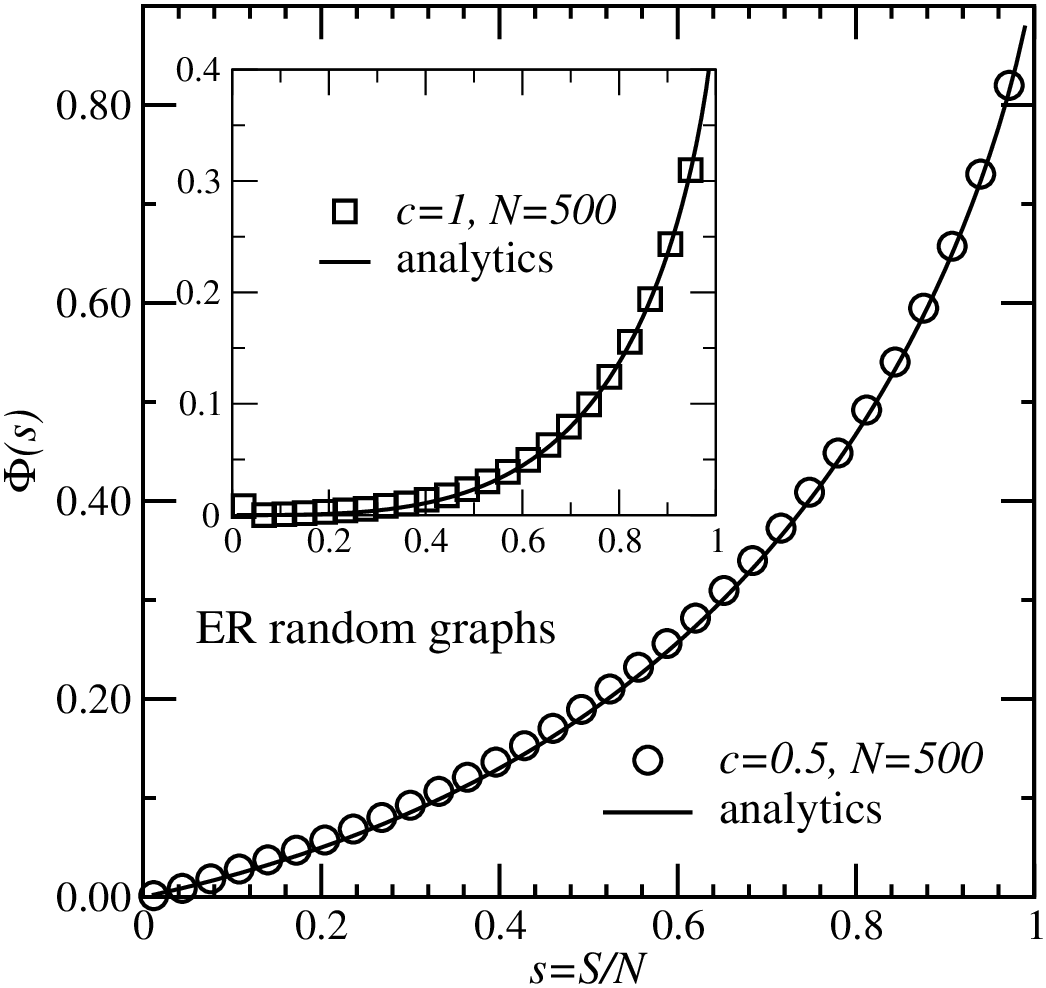}
  \caption{
    \label{fig:rateER0.5_1.0}
Large-deviation rate function $\Phi(s)$ 
of ER random graphs with 
average connectivity $c=0.5<c_{\rm c}$, $N=500$ (symbols). The
line displays the analytical result from Eq. (\ref{eq:rate_fct_analytic}).
The inset shows the same for the case $c=1.0=c_{\rm c}$.
}
\end{figure}

In Fig.\ \ref{fig:rateER0.5_1.0} the empirical rate function 
\begin{equation}
\Phi(s)\equiv-\frac{1}{N} \log P(s)
\end{equation}
for $c=0.5$ is displayed, corresponding
to the distribution shown in Fig.\ \ref{fig:equil_distr}.
Note that by just stating the analytical asymptotic rate function 
$\Phi_{\rm ER}(s,c)$, the corresponding distribution $P(s)$ is not 
normalised. Hence, for comparison, $\Phi(s)$ is shifted for all values of 
the connectivity $c$
such that it is zero at its minimum value, like $\Phi_{\rm ER}(s, c)$.

The numerical data agrees very well with the analytic result. 
Only in the region of intermediate cluster sizes, a small systematic deviation
is visible, which is likely to be a finite-size effect. Given that
for the numerical simulations only graphs with $N=500$ nodes were treated,
the agreement with the $N\to\infty$ leading-order analytical result is 
remarkable.

The resulting rate function right at the percolation transition 
$c=c_{\rm c}=1$ is
shown in the inset of Fig.\ \ref{fig:rateER0.5_1.0}. Qualitatively,
the result is very similar to the non-percolating case $c=0.5$, except that
the distribution is much broader, corresponding to smaller values
of the rate function. Again, the agreement with the analytical result
is very good, except for the data close to the origin $s=0$: The numerical
results exhibit a minimum near $s=0.05$, while the analytical result
exhibits its minimum naturally at $s=0$. This is clearly due to the
finite size of the numerical samples.

\begin{figure}[ht]
  \centering
  \includegraphics[clip,width=0.45\textwidth]{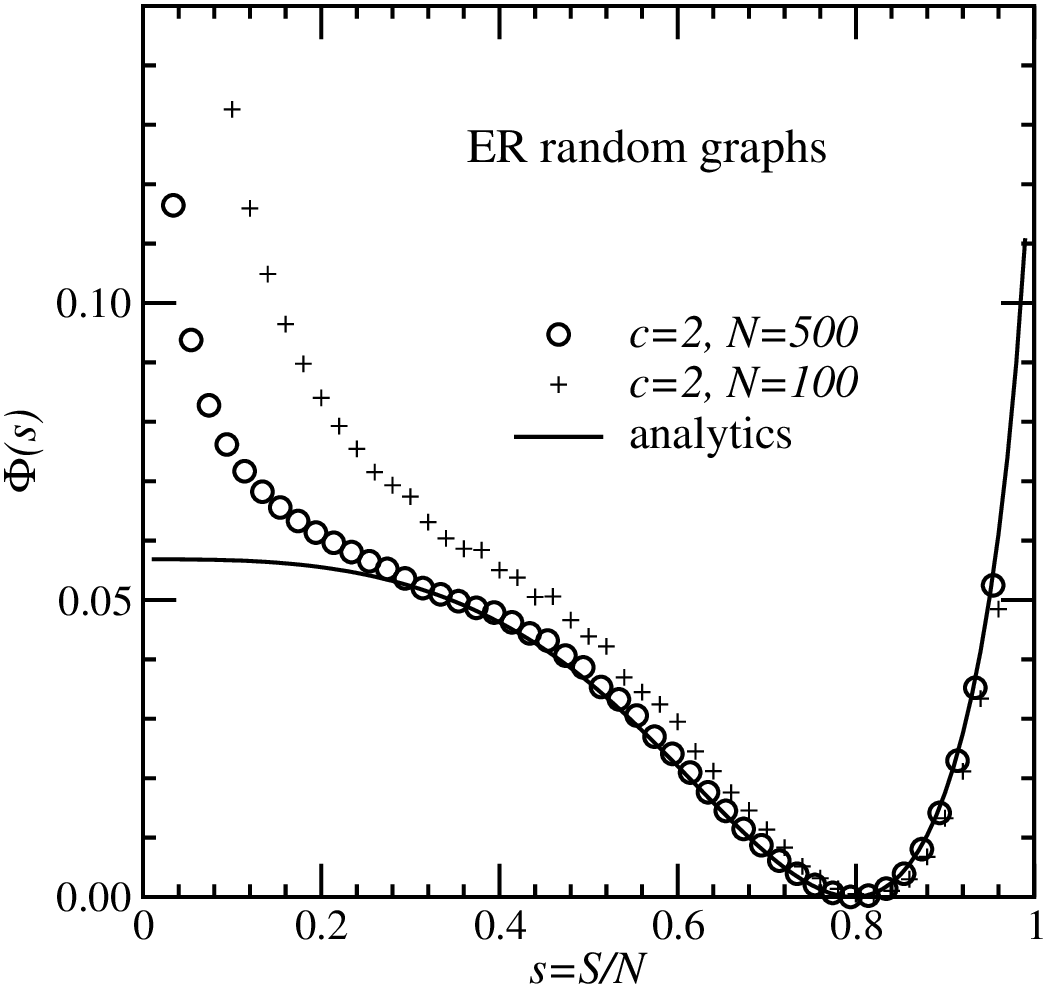}
  \caption{
    \label{fig:rateER2.0}
Large-deviation rate function $\Phi(s)$ 
of ER random graphs with 
average connectivity $c=2>c_{\rm c}$, $N=100$ and $N=500$ (symbols). The
line displays the analytical result from Eq. (\ref{eq:rate_fct_analytic}).
}
\end{figure}

The case of the percolating regime (connectivity $c=2$), is
displayed in Fig.\ \ref{fig:rateER2.0}. The rate function
 exhibits a minimum at a finite value of $s$,
corresponding to the finite average fraction of nodes contained 
in the largest component.
The behaviour of the rate function
is more interesting compared to $c\le c_{\rm c}$, because $\Phi_{\rm ER}(s)$
grows strongly near its minimum, but for $s\to 0$ it levels off horizontally. 
For  most of
the support of the distribution, 
the numerical data for $N=500$ agrees again very well
with the analytic result. Nevertheless, for $s\to 0$, strong deviations
become visible because the numerical rate function $\Phi(s)$ grows strongly
as $s\to 0$.
By comparing with the result for a smaller system, $N=100$, where
this deviation is even larger, it becomes clear that this is a 
finite-size effect, corresponding to the non-leading corrections, which
will disappear for $N\to\infty$.

\begin{figure}[ht]
  \centering
  \includegraphics[clip,width=0.45\textwidth]{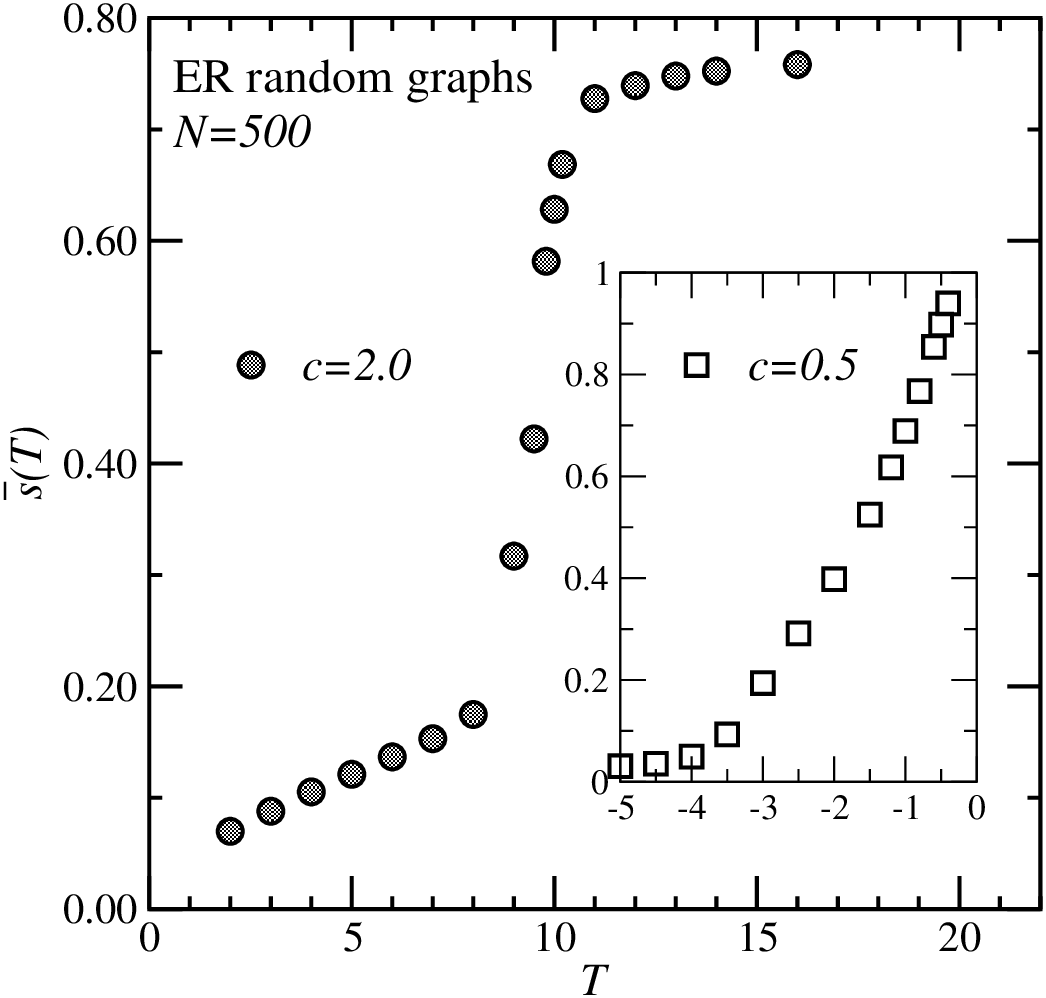}
  \caption{
    \label{fig:STER}
Average relative size $\overline{s}$ of the largest component as a function
of artificial temperature $T$ for ER random graphs in the percolating regime
($c=2$) and (inset) in the non-percolating regime ($c=0.5$).
}
\end{figure}

Although the parameter $T$ is mainly used as a way to address different
parts of the distribution, the temperature dependence of $S$,
which is shown in Fig.\ \ref{fig:STER}, exhibits
an interesting behaviour on its own in the percolating regime.
When studying the average size $\overline{s}$ as a function of
temperature, a strong increase around temperature $T=9.5$ becomes visible,
which may correspond to a kind of phase transition, see below.

On the other hand, in the non-percolating regime $c<1$, the average
size of the largest component is rather smooth, as shown in the 
inset of Fig.\ \ref{fig:STER}. Note that here negative artificial temperatures
are used, because the peak of the distribution $P(S)$ is close to $S=0$,
in contrast to the percolating regime $c>1$, where the peak of $P(S)$
is at finite values.

\begin{figure}[ht]
  \centering
  \includegraphics[clip,width=0.45\textwidth]{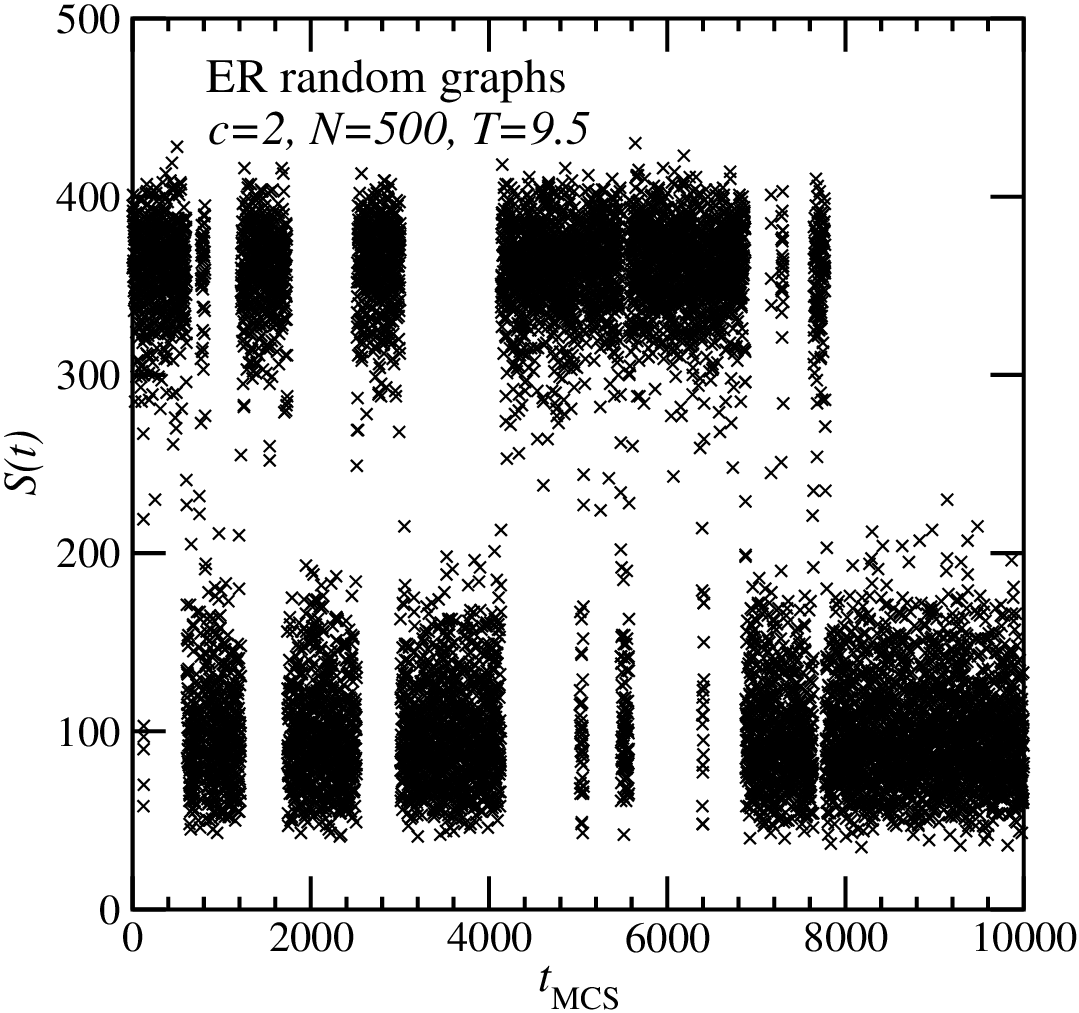}
  \caption{
    \label{fig:PtERT}
Time-series for the size $S$ of largest cluster as function of the number
$t_{\rm MCS}$
of MC sweeps for ER random graphs
($c=2$, $N=500$) at artificial temperature $T=9.5$.
}
\end{figure}

In Fig.\ \ref{fig:PtERT} the size $S$ of the largest component is shown as a 
function of the number of MC sweeps for a temperature $T=9.5$, which is
located in the
regime of the assumed transition.  One can see that $S(t)$ fluctuates
quickly between two sets of typical sizes, which shows also that the
data is well equilibrated. In particular, values around $S=350$ and
around $S=100$ are more frequent than intermediate values.

\begin{figure}[ht]
  \centering
  \includegraphics[clip,width=0.45\textwidth]{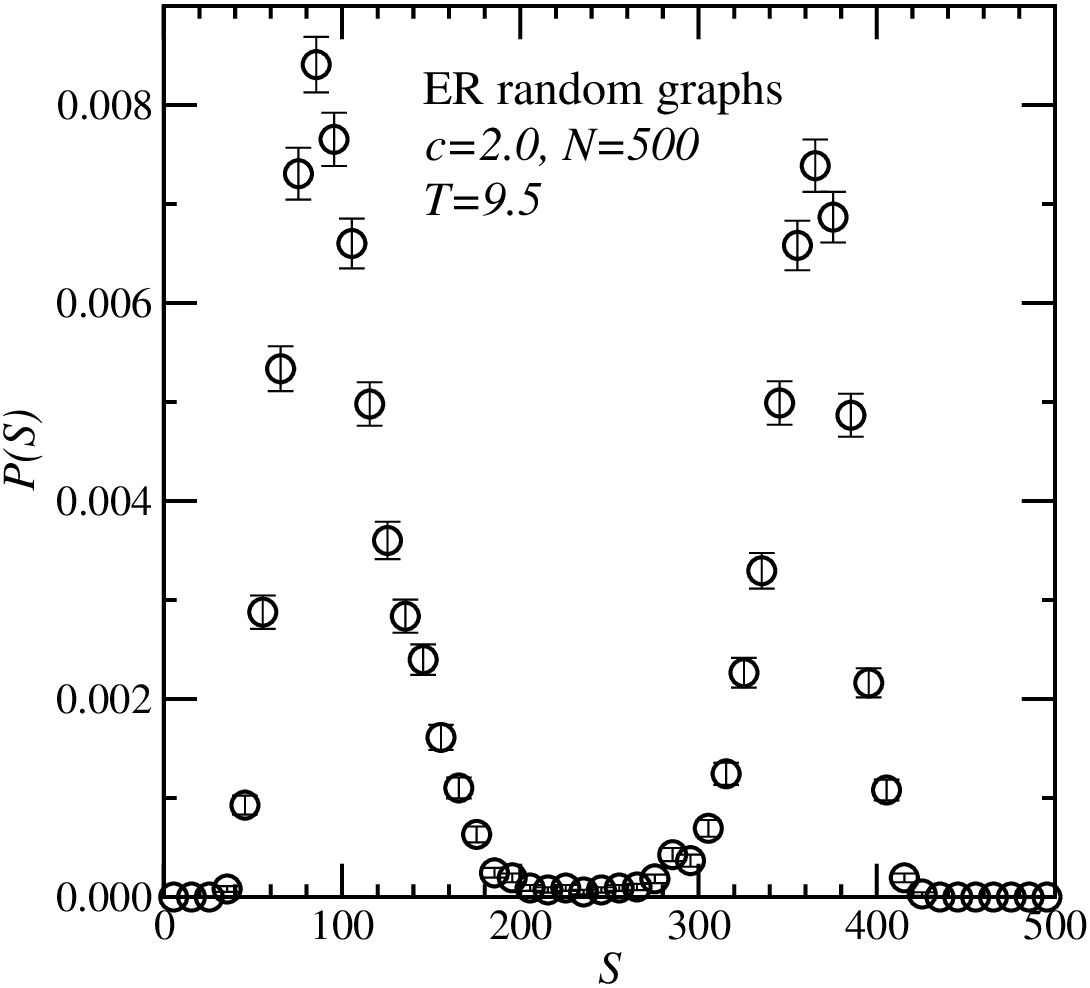}
  \caption{
    \label{fig:distrERPsT}
Distribution (normalised such that the integral is one)
of the size of the largest component for ER random graphs 
($c=2$, $N=500$) at artificial temperature $T=9.5$.
}
\end{figure}

This result is made more quantitative by studying 
 the resulting distribution of the sizes of the largest
component, see Fig.\ \ref{fig:distrERPsT}.
A two-peak structure can
be observed, which indicates that indeed a transition between small and
large components of first-order
type is present in the percolating regime $c>1$.
 Similar first-order transitions have been observed for biased
simulations of the one-dimensional Ising model \cite{jack2010,jack2010b}. 
The first-order nature of this transition is a reason that obtaining
$\Phi(s)$ becomes harder for large system sizes, because the time for
tunnelling between the two sets of values grows quickly. Furthermore, even worse,
the number of observed configurations having a value of $S$
 which is located between the peaks decreases strongly, such that for large
intervals no data can be collected at all, already for $N=1000$.
 Hence, $P(S)$ cannot be sampled for such system sizes 
on its complete support, because the different parts of the distribution 
cannot be ``glued'' together.

A detailed study of this phase transition is beyond the scope of this
work, in particular because its physical relevance is yet not fully clear.

\subsection{Two-dimensional percolation}

\begin{figure}[t!]
  \centering
  \includegraphics[clip,width=0.45\textwidth]{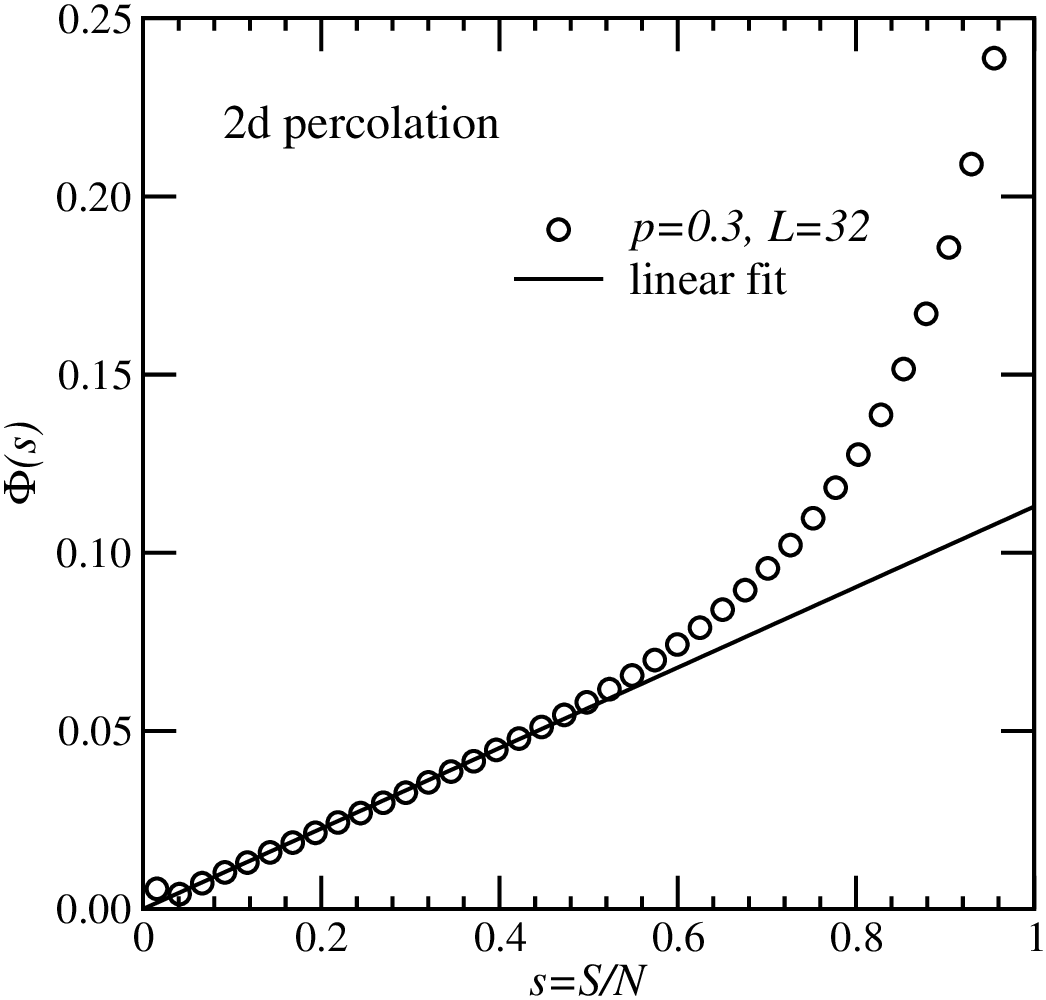}
  \caption{
    \label{fig:rate2d0.3}
Large-deviation rate function $\Phi(s)$ 
of two-dimensional (2d) bond percolation with 
occupation probability $p=0.3<p_{\rm c}$, $L=32$ (symbols). For
small relative cluster sizes $s$, the rate function behaves linearly
(the line displays a linear function with slope $0.1125$, obtained
from fitting a linear function to the data in the range $s\in[0,0.5]$).
}
\end{figure}

Again the discussion of the rate function for the distribution of the
relative size $s=S/N$ of the largest component is started by considering
the non-percolating regime. 
The rate function for $p=0.3<p_{\rm c}$ ($L=32$) is shown in
Fig.\ \ref{fig:rate2d0.3}. In principle it looks similar to the ER
case displayed in Fig.\ \ref{fig:rateER0.5_1.0}. A striking difference is that
it exhibits a large region where
it behaves linearly, basically for half of the support, while it grows
stronger for $s>0.5$.
 For $L=16$ (not shown), 
the same result was found. This means, the finite-size effects are small
as in the ER case. This also indicates that the shape of the rate function
should basically remain the same for $L\to\infty$.
 In particular, it appears likely that for $L\to\infty$, 
$\Phi(s)$ will consist of a linear part for small values of $s$ and it will
grow stronger for $s\to 1$.

\begin{figure}[ht]
  \centering
  \includegraphics[clip,width=0.45\textwidth]{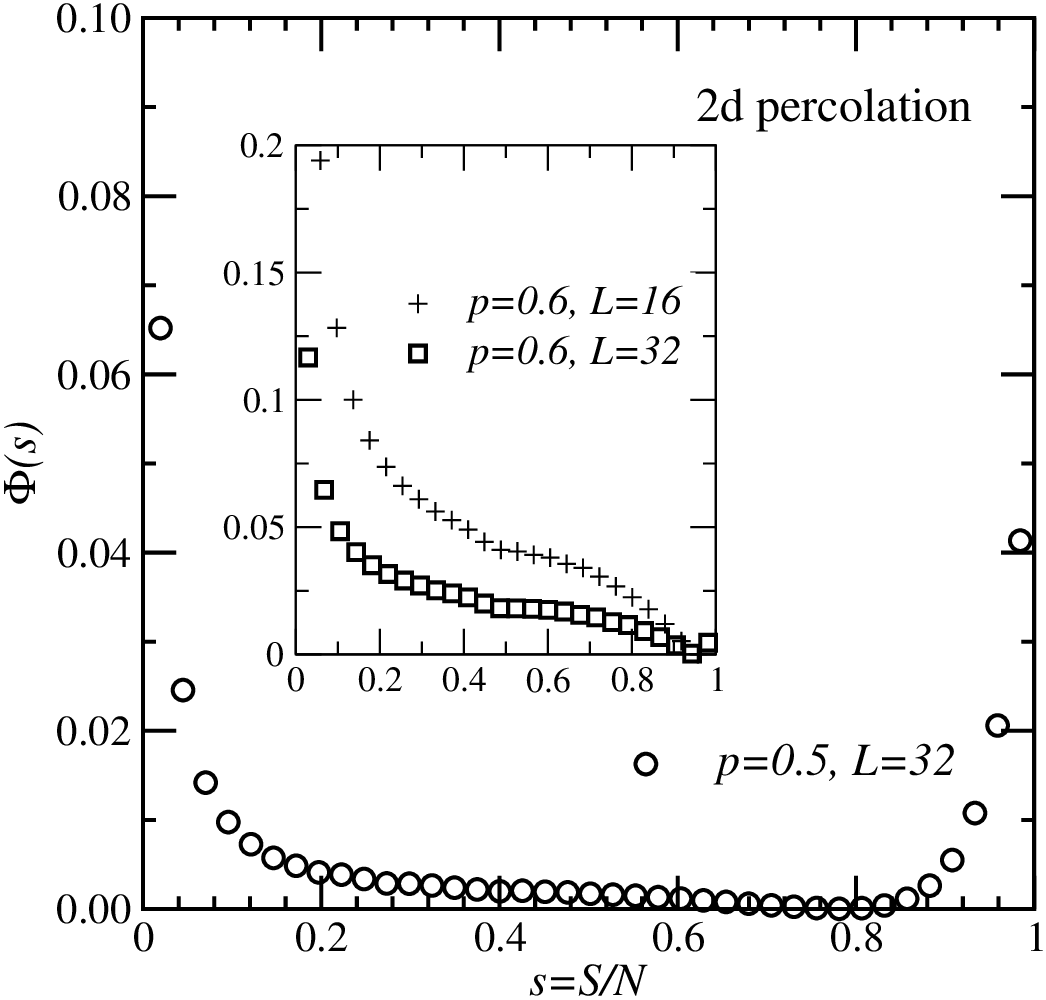}
  \caption{
    \label{fig:rate2d0.5_0.6}
Large-deviation rate function $\Phi(s)$ 
of two-dimensional (2d) bond percolation with 
occupation probability $p=0.5=p_{\rm c}$, $L=32$ and (inset) $p=0.6$, $L=16$/$L=32$.
}
\end{figure}

In contrast, the large-deviation rate function right at the percolation
transition $p=p_{\rm c}=0.5$ ($L=32$), see Fig.\ \ref{fig:rate2d0.5_0.6}, 
looks  very different from the ER case: It exhibits a
minimum at a relative cluster size of about $s=S/N=0.78$,
see main plot in Fig.\ \ref{fig:rate2d0.5_0.6}. This is very large
compared to the ER case, but only a finite-size effect: For example for
$L=512$ (additional simple sampling simulations, not shown), 
the minimum of the rate function has already moved to a smaller value 
of $s\approx 0.5$ and for $p=0.49$, just before the percolation transition,
the most likely relative size of a cluster
is $s=0.1$. Hence, the finite-size effects, i.e., corrections
to the large-deviation rate function, are close to $p_c$ much stronger
for the finite-dimensional case compared to ER random graphs.  

The case of the percolating regime is displayed in the inset of
Fig.\ \ref{fig:rate2d0.5_0.6}. Here again, the result looks very similar
to the ER case, except that the magnitude of the rate
function is somehow smaller. Hence, it appears likely that the shape
of the rate function for the 2d percolation problem in the
limit $L\to\infty$ is very similar to the ER case, i.e.,
it may level off horizontally for $s\to 0$ at a finite value and the strong
increase found for $L=32$ is again a finite size effect. This is
confirmed by the result for $L=16$, which exhibits a much stronger increase
for $s\to 0$ compared to the $L=32$ result, as in the case of ER random graphs.

\begin{figure}[ht]
  \centering
  \includegraphics[clip,width=0.45\textwidth]{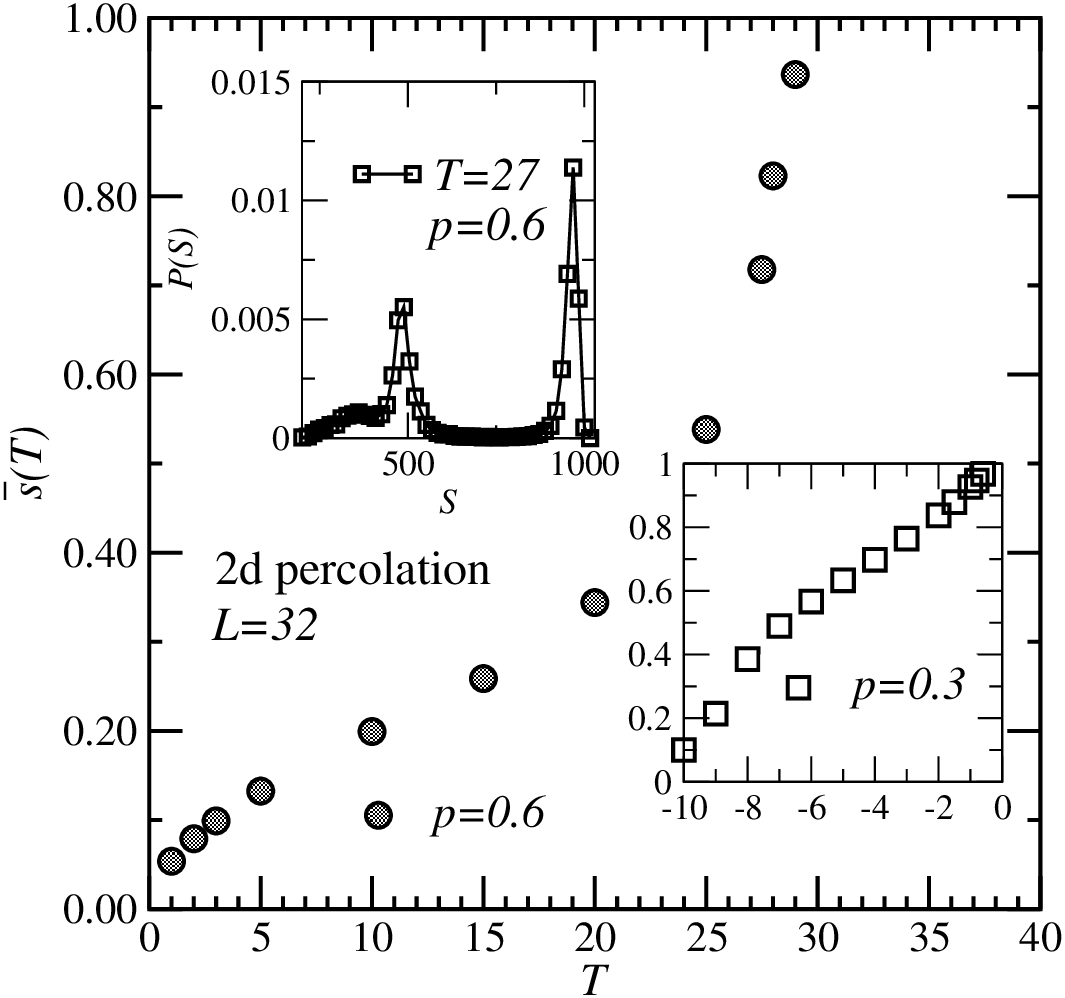}
  \caption{
    \label{fig:STdistrPerc}
Average relative size $\overline{s}$ of the largest component as a function
of artificial temperature $T$ for 2d percolation ($L=32$) 
in the percolating regime
($p=0.6$) and (right inset) in the non-percolating regime ($p=0.6$).
The left inset shows the distribution 
(normalised such that the integral is one)
of the size of the largest component  
($p=6$) at artificial temperature $T=27$ (line is guide to the eyes only).
}
\end{figure}

In Fig.\ \ref{fig:STdistrPerc} the result for the behaviour
of $S$ as a function of the artificial temperature is shown. In the 
non-percolating regime ($p=0.3$), see right 
inset, the average value $\overline s$ behaves 
very regularly as a function of the temperature, no sign of a transition
is visible. On the other hand,
inside the percolation regime ($p=0.6$), see main plot, $\overline{s}(T)$ 
exhibits a strong increase around $T=27$. The 
distribution of $S$ at $T=27$ exhibits a strong bimodal signature, which
indicates that indeed a phase transition of first-order type takes place.
In summary, the temperature dependence of the size of the largest
component is very similar to the results obtained for the ER random graphs.

\section{Summary and outlook}

By using an artificial Boltzmann ensemble characterised by an artificial
temperature $T$, the distributions of the
size of the largest component for ER randoms graphs with finite
connectivity $c$ and for 2d
percolation have been studied in this work. For not too large
system sizes, the distributions can be calculated numerically
over the full support, giving access to very small probabilities
such as $10^{-180}$.

For the ER case,
the numerical results for the large-deviation rate function $\Phi(s)$, obtained 
for rather small graphs of size $N=500$, agree very
well with analytical results obtained previously for the leading behaviour
in the limit $N\to\infty$. This proves the usefulness of the numerical
approach, which has been applied previously to models where no
complete comparison between numerical data and exact analytic results
have been performed. 
The main findings are that below and at the percolation transition, $\Phi(s)$
exhibits a minimum at $s=0$ and rises monotonously  for $s\to 1$.  Inside
the percolating regime, $\Phi(s)$ exhibits a minimum, grows quickly
around this minimum and levels off horizontally for $s\to 0$. The finite-size
corrections are usually small, except for the percolating regime in
an extended region near $s=0$. Furthermore, when studying the average value
$\overline s$ as a function of temperature, a transition of first-order
type is found between a phase where $S$ is untypically small to a phase where
$S$ is large.

For the 2d percolation problem, where no analytic results are available,
basically the same results are found: the shape of the large-deviation
rate functions below, at, and above the percolation threshold
are qualitatively the same, except that the finite-size corrections
at the percolation threshold appear to be larger compared to the ER results.
Also the behaviour of the largest-component size  as a function of the
temperature seems to be similar, in particular exhibiting a first-order type
transition for the percolating regime.

Since the comparison with the exact results for the ER random
graphs indicates the usefulness
of this approach to study large-deviation properties of random graphs, 
it appears promising to consider many other properties of different ensembles
of random graphs in the same way. For example, it would
be interesting to obtain the distribution of the diameter of ER random graphs,
where only for $c<1$ there is an analytic result available. Corresponding
simulations are currently performed by the author of this work.

\section*{Acknowledgements}
The author thanks Oliver Melchert for critically reading the manuscript.
 The simulations  were partially performed at the GOLEM I cluster for 
scientific computing at the University of Oldenburg (Germany). 

\bibliographystyle{epj}
\bibliography{alex_refs,remarks}

\end{document}